\documentclass[epj]{svjour}

\usepackage{graphics}

\usepackage{epsfig}

\def\lsim{\raise0.3ex\hbox{$<$\kern-0.75em\raise-1.1ex\hbox{$\sim$}}}
\def\gsim{\raise0.3ex\hbox{$>$\kern-0.75em\raise-1.1ex\hbox{$\sim$}}}

\begin{document}

\title{Running coupling of 2-flavor QCD at zero and finite temperature}

\author{Olaf Kaczmarek\inst{1} \and Felix Zantow\inst{2}} \institute{Fakult\"at
  f\"ur Physik, Bielefeld University, D-33615 Bielefeld, Germany \and Physics
  Department, Brookhaven Natl. Laboratory, Upton, NY-11973, USA }

\date{\today}

\abstract{We present lattice studies of the running coupling in $2$-flavor QCD.
  The coupling at zero temperature ($T=0$) is extracted from Wilson loops while
  the coupling at finite temperature ($T\neq0$) is determined from Polyakov
  loop correlation functions. }

\PACS{{11.15.Ha,}{ 11.10.Wx,}{ 12.38.Mh,}{ 25.75.Nq}}
                                                                                
\maketitle

\section{Running couplings}
The QCD coupling plays an important role at zero temperature and, in
particular, at finite temperature in todays discussion of possible signals for
the quark gluon plasma formation in heavy ion experiments
\cite{Brambilla,Shuryak,Brown}. We calculate running couplings from lattice
studies of the Wilson loop ($T=0$) (from \cite{Peikert}) and Polyakov loop
correlation functions ($T\neq0$) in $2$-flavor QCD ($N_f=2$) using an improved
staggered fermion action with quark mass $m/T=0.4$ (corresponding to $ma=0.1$)
\cite{Allton}. Any further details on this study can be found in
\cite{Peikert,Kacze03a,Kacze05/1,Kacze05/2}. Similar studies in quenched QCD are reported
in Refs.~\cite{Necco,Kacze04,Kacze02}. First experiences with the running
coupling at finite temperature in $3$-flavor QCD are reported in \cite{Peter}.

\subsection{Heavy quark potential at $T=0$}
\begin{figure}[tbp]
  \epsfig{file=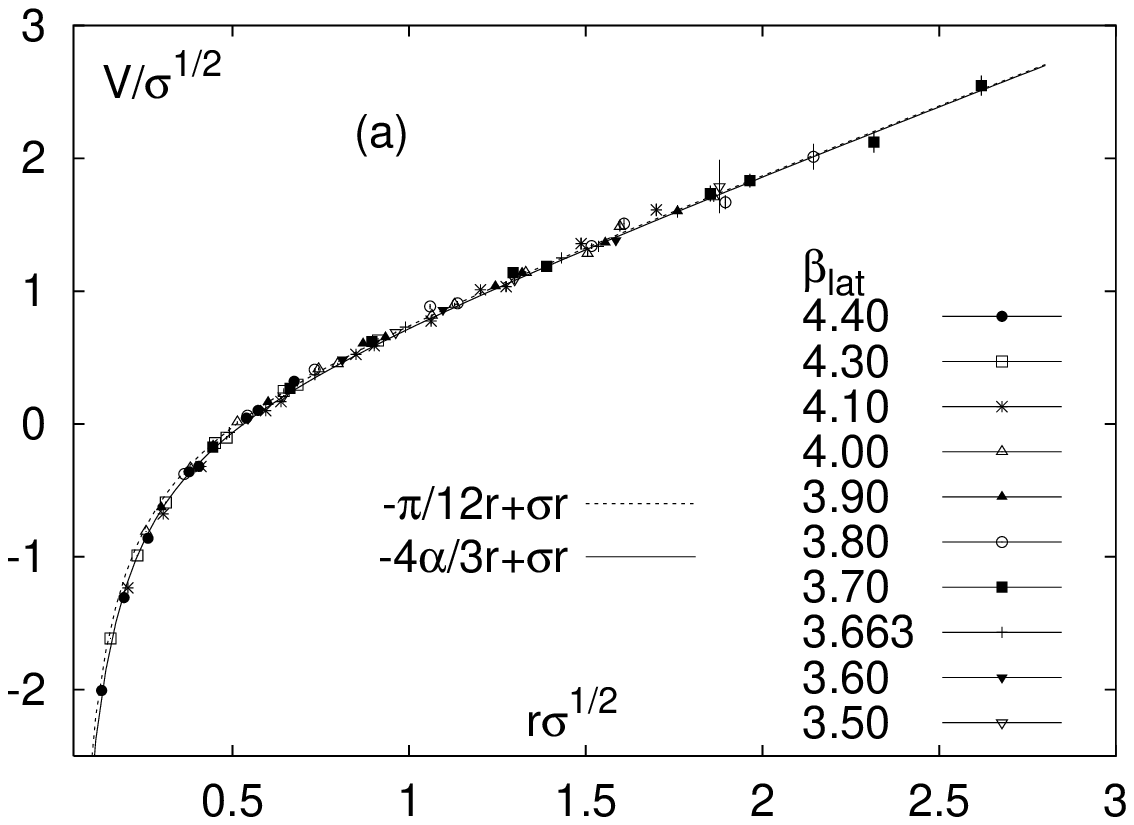,width=9.0cm}
  \epsfig{file=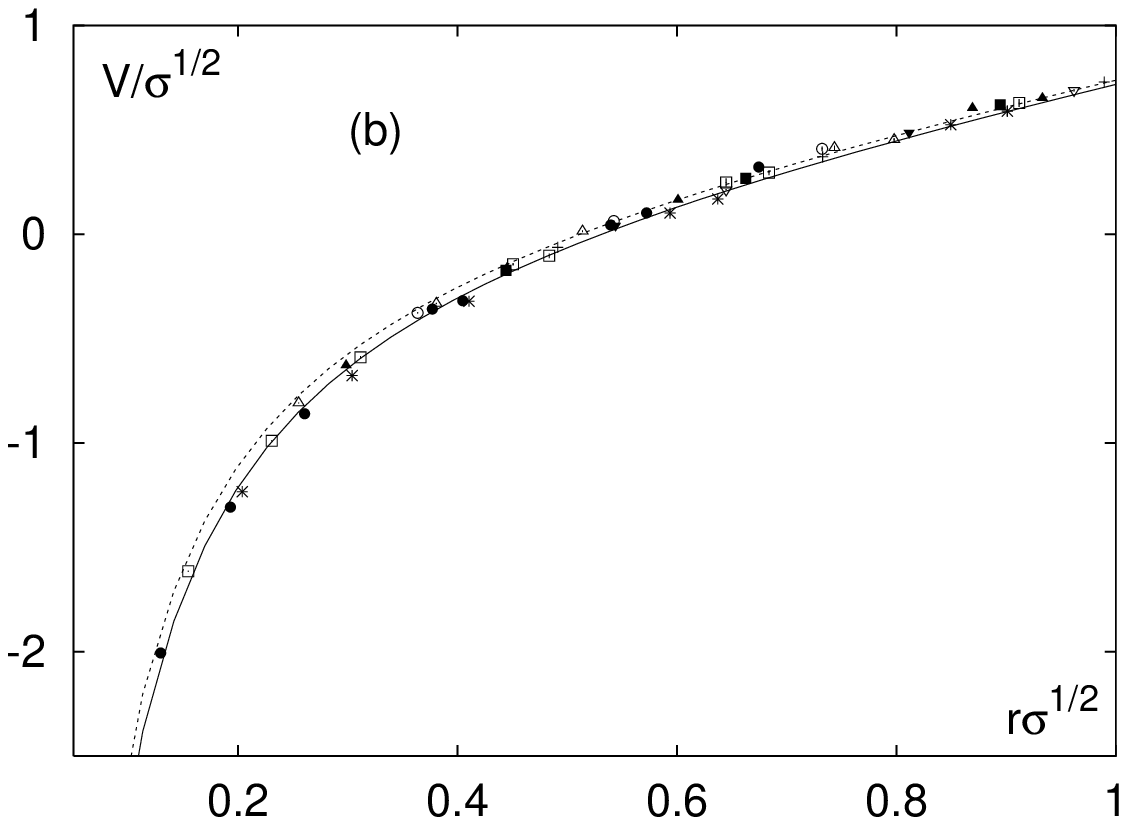,width=9.0cm}
\caption{
  (a) The heavy quark potential at $T=0$ from \cite{Peikert} obtained from
  2-flavor QCD lattice simulations with quark masses $ma=0.1$ for different
  values of the lattice coupling $\beta$. Fig.~2b shows an enlargement of the
  short distance regime. The data are matched to the bosonic string potential
  (dashed line) at large distances. Included is also the fit to the Cornell
  form (solid line) given in Eq.~\ref{t=0ansatz}.  }
\label{peik}
\end{figure}

For the determination of the heavy quark potential at zero temperature, $V(r)$,
we have used the measurements of large smeared Wilson loops given in
\cite{Peikert} ($N_f$=2 and $ma=0.1$). To eliminate the divergent self-energy
contributions we matched these data for all $\beta$-values (different
$\beta$-values correspond to different values of the lattice spacing $a$) at
large distances to the bosonic string potential,
\begin{eqnarray}
V(r) &=& - \frac{\pi}{12}\frac{1}{r} + \sigma r \;\equiv\;-\frac{4}{3}\frac{\alpha_{{str}}}{r}+\sigma r\;,
\label{string-cornell}
\end{eqnarray}
where we already have separated the Casimir factor so that
$\alpha_{{str}}\equiv\pi/16$. In Fig.~\ref{peik}a,b we show our results
together with the heavy quark potential from the string picture (dashed line).
One can see that the data are well described by Eq.~\ref{string-cornell} at
large distances, {\em i.e.}  $r\sqrt{\sigma}\;\gsim\;0.8$, corresponding to
$r\;\gsim\;0.4$ fm. At these distances we see no major difference between the
2-flavor QCD potential obtained from Wilson loops and the quenched QCD
potential which is well described by the string model already for
$r\;\gsim\;0.4$ fm \cite{Necco,Luscher}. In fact, we also do not see any signal
for string breaking in the zero temperature QCD heavy quark potential. This is
to some extend due to the fact that the Wilson loop operator used here for the
calculation of the $T=0$ potential has only small overlap with states where
string breaking occurs \cite{Bernard}. Moreover, the distances for which we
analyze the data for the QCD potential are all below $r\;\lsim\;1.5$ fm at
which string breaking is expected to set in at zero temperature.

\subsection{The coupling at $T=0$}\label{couplt=0}
Deviations from the string model and from the pure gauge potential, however,
are clearly expected to become apparent in the 2-flavor QCD potential at small
distances and may already be seen from the short distance part in
Fig.~\ref{peik}. These deviations are expected to arise from an asymptotic
weakening of the QCD coupling, {\em i.e.} $\alpha=\alpha(r)$, and also is to
some extent due to the effect of including dynamical quarks ($N_f\neq0$), {\em
  i.e.} from leading order perturbation theory one expects
\begin{eqnarray}
\alpha(r) \simeq \frac{1}{8\pi} \frac{1}{\beta_0 \log \left(1/(r \Lambda_{{ QCD}})\right)}\;,
\label{runningcoupling}
\end{eqnarray}
with $\beta_0 = (33-2N_f)/(48 \pi^2)$ where $N_f$ is the number of flavors. The
data in Fig.~\ref{peik}b show a slightly steeper slope at distances below
$r\sqrt{\sigma}\simeq0.5$ compared to the pure gauge potential given in
Ref.~\cite{Necco} indicating that the QCD coupling gets stronger in the entire
distance range analyzed here when including dynamical quarks. To include the
effect of a stronger Coulombic part in the QCD potential we test the Cornell
parameterization,
\begin{eqnarray}
\frac{V(r)}{\sqrt{\sigma}} = -\frac{4}{3}\frac{\alpha}{r\sqrt{\sigma}} + r \sqrt{\sigma}
\label{t=0ansatz}\;,
\end{eqnarray}
with a free parameter $\alpha$. From a best fit analysis of Eq.~\ref{t=0ansatz}
to the data ranging from $0.2\;\lsim\;r\sqrt{\sigma}\;\lsim\;2.6$ we find
$\alpha=0.212(3)$.  This already may indicate that the logarithmic weakening of
the coupling with decreasing distance will not too strongly influence the
properties of the QCD potential at these distances, {\em i.e.} at
$r\;\gsim\;0.1$ fm. However, the value of $\alpha$ is moderately larger than
$\alpha_{{str}}\;\simeq\;0.196$ introduced above. To compare the relative size
of $\alpha$ in full QCD to $\alpha$ calculated in the quenched theory we again
have performed a best fit analysis of the quenched zero temperature potential
given in \cite{Necco} using the ansatz given in Eq.~\ref{t=0ansatz} and a
similar distance range. Here we find $\alpha_{{quenched}} = 0.195(1)$ which is
again smaller than the value for the QCD coupling but quite comparable to
$\alpha_{{str}}$.

When approaching the short distance perturbative re-\\ gime a Cornell ansatz
will overestimate the value of the coupling due to the perturbative logarithmic
weakening of the latter, $\alpha_{{QCD}}=\alpha_{{QCD}}(r)$.  To analyze the
short distance properties of the QCD potential and the coupling in more detail,
{\em i.e.} at $r\;\lsim\;0.4$ fm, and to firmly establish here the onset of its
perturbative weakening with decreasing distance, it is customary to do so using
non-perturbative definitions of running couplings. Following recent discussions
on the running of the QCD coupling \cite{Necco,Schroeder,Necco2}, it appears
most convenient to study the QCD force, {\em i.e.} $dV(r)/dr$, rather than the
QCD potential. In this case one defines the QCD coupling in the so-called
$qq$-scheme,
\begin{eqnarray}
\alpha_{qq}(r)&\equiv&\frac{3}{4}r^2\frac{dV(r)}{dr}\;.
\label{alp_qq}
\end{eqnarray} 
In this scheme any undetermined constant contribution to the heavy quark
potential cancels out. Moreover, the large distance, non-perturbative
confinement contribution to $\alpha_{qq}(r)$ is positive and allows for a
smooth matching of the perturbative short distance coupling to the
non-perturbative large distance confinement signal.

\begin{figure}[tbp]
  \epsfig{file=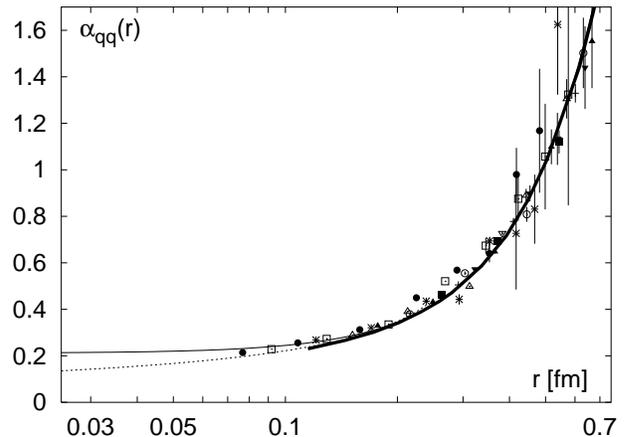,width=9.0cm}
\caption{
  The short distance part of the running coupling $\alpha_{qq}(r)$ in 2-flavor
  QCD at zero temperature defined in Eq.~\ref{alp_qq} as function of the
  distance $r$ (in physical units). The symbols for the different
  $\beta$-values are chosen as indicated in Fig.~\ref{peik}a. The lines are
  discussed in the text.  }
\label{peiks}
\end{figure}
Our results for $\alpha_{qq}(r)$ as a function of distance in physical units
for 2-flavor QCD are summarized in Fig.~\ref{peiks}. The symbols for the
different $\beta$-values are chosen as in Fig.~\ref{peik}a. We again show in
that figure the corresponding line for the Cornell fit (solid line). At large
distances, $r\;\gsim\;0.4$ fm, the data clearly mimic the non-perturbative
confinement part of the QCD force, $\alpha_{qq}(r)\simeq3r^2\sigma/4$. We also
compare our data to the recent high statistics calculation in pure gauge theory
(thick solid line). These data are available for $r\;\gsim\;0.1$ fm and within
the statistics of the QCD data no significant differences could be identified
between the QCD and pure gauge data for $r\;\gsim\;0.4$ fm. At smaller
distances ($r\;\lsim\;0.4$ fm), however, the data show some enhancement
compared to the coupling in quenched QCD. The data below $0.1$ fm, moreover,
fall below the large distance Cornell fit. This may indicate the logarithmic
weakening of the coupling. At smaller distances than $0.1$ fm we therefore
expect the QCD potential to be influenced by the weakening of the coupling and
$\alpha_{qq}(r)$ will approach values clearly smaller than $\alpha$ deduced
from the Cornell ansatz. Unfortunately we can, at present, not go to smaller
distances to clearly demonstrate this behavior with our data in 2-flavor QCD.
Moreover, at small distances cut-off effects may also influence our analysis of
the coupling and more detailed studies are required here. In earlier studies of
the coupling in pure gauge theory \cite{Necco,Kacze04,Necco2} it has, however,
been shown that the perturbative logarithmic weakening becomes already
important at distances smaller than $0.2$ fm and contact with perturbation
theory could be established.

\subsection{The running coupling at $T\neq0$}\label{couplatt}
We extend here our studies of the coupling at zero temperature to finite
temperature below and above deconfinement following the conceptual approach
given in \cite{Kacze04,Kacze02}. In this case the appropriate observable is the
color singlet quark anti-quark free energy and its derivative. We use the
perturbative short and large distance relation from one gluon exchange
\cite{Nadkarni,Nadkarni2,McLerran}, {\em i.e.} in the limit
$r\Lambda_{{QCD}}\ll1$ zero temperature perturbation theory suggests
\begin{eqnarray}
F_1(r,T)&\simeq&-\frac{4}{3}\frac{\alpha(r)}{r}\;,\label{alp_rT1}
\end{eqnarray}
while high temperature perturbation theory, {\em i.e.} $rT\gg1$ and $T$ well
above $T_c$, yields
\begin{eqnarray}
F_1(r,T)&\simeq&-\frac{4}{3}\frac{\alpha(T)}{r}e^{-m_D(T)r}\;.\label{alp_rT2}
\end{eqnarray}
In both relations we have neglected any constant contributions to the free
energies which, in particular, at high temperatures will dominate the large
distance behavior of the free energies. Moreover, we already anticipated here
the running of the couplings with the expected dominant scales $r$ and $T$ in
both limits. At finite temperature we define the running coupling in analogy to
$T=0$ as,
\begin{eqnarray}
  \alpha_{qq}(r,T)&\equiv&\frac{3}{4}r^2
  \frac{dF_1(r,T)}{dr}\;.\label{alp_rT}
\end{eqnarray}  
With this definition any undetermined constant contributions to the free
energies are eliminated and the coupling defined here at finite temperature
will recover the coupling at zero temperature defined in (\ref{alp_qq}) in the
limit of small distances. Therefore $\alpha_{qq}(r,T)$ will show the (zero
temperature) weakening in the short distance perturbative regime. In the large
distance limit, however, the coupling will be dominated by Eq.~\ref{alp_rT2}
and will again be suppressed by color screening,
$\alpha_{qq}(r,T)\sim\exp(-m_D(T)r)$, $rT\gg1$. It thus will exhibit a maximum
at some intermediate distance.

\begin{figure}[tbp]
  \epsfig{file=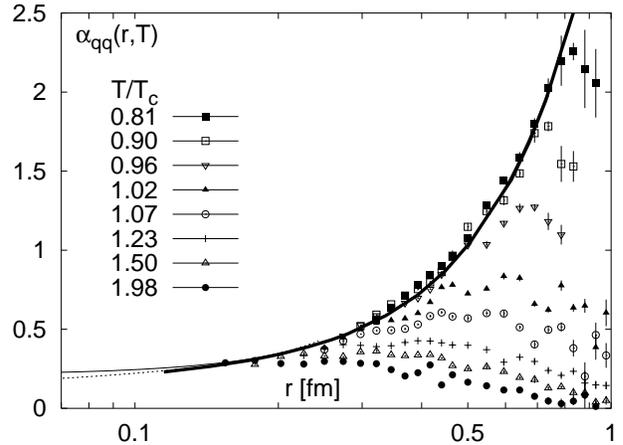,width=9.0cm}
\caption{
  The running coupling in the $qq$-scheme defined in Eq.~\ref{alp_rT}
  calculated from derivatives of the color singlet free energies with respect
  to $r$ at several temperatures as function of distance below and above
  deconfinement. We also show the corresponding coupling at zero temperature
  (solid line) from Eq.~\ref{t=0ansatz} and compare the results again to the
  results in pure gauge theory (thick solid and dashed lines)
  \cite{Necco,Necco2}.  }
\label{couplt}
\end{figure}
Lattice results for $\alpha_{qq}(r,T)$ calculated in this way are shown in
Fig.~\ref{couplt} and are compared to the coupling at zero temperature
discussed already in Sec.~\ref{couplt=0}. Here the thin solid line corresponds
to the coupling in the Cornell ansatz given in Eq.~\ref{t=0ansatz}. We again
show in this figure the results from $SU(3)$-lattice (thick line) and
perturbative (dashed line) calculations at zero temperature from
\cite{Necco,Necco2}. The strong $r$-dependence of the running coupling near
$T_c$ observed already in pure gauge theory \cite{Kacze04,Kacze02} is also
visible in 2-flavor QCD.  Although our data for 2-flavor QCD do not allow for a
detailed quantitative analysis of the running coupling at smaller distances,
the qualitative behavior is in quite good agreement with the recent quenched
results. At large distances the running coupling shows a strong temperature
dependence which sets in at shorter distances with increasing temperature. For
small temperatures, $T\;\lsim\;1.02T_c$, the coupling $\alpha_{qq}(r,T)$
already coincides with $\alpha_{qq}(r)$ at distance $r\;\simeq\;0.4$ fm and
clearly mimics here also the confinement part of $\alpha_{qq}(r)$. This is also
apparent in quenched QCD \cite{Kacze04}. Remnants of the confinement part of
the QCD force may survive the deconfinement transition. A clear separation of
the different effects usually described by the concepts of color screening
($T\;\gsim\;T_c$) and effects commonly described by the concept of
string-breaking ($T\;\lsim\;T_c$) is difficult to establish at temperatures in
the close vicinity of the confinement deconfinement cross over.

\begin{figure}[tbp]
  \epsfig{file=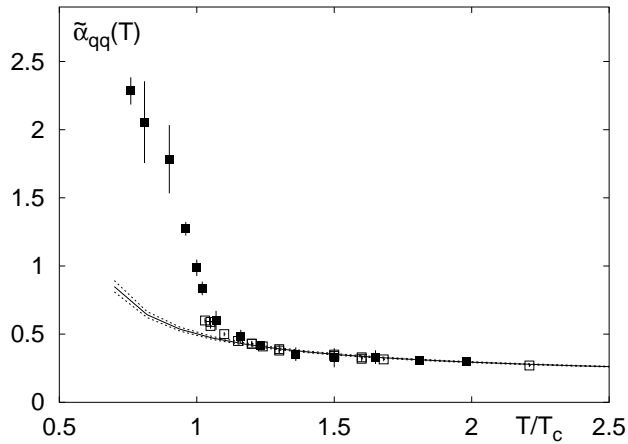,width=9.0cm}
\caption{
  The size of the maximum, $\tilde{\alpha}_{qq}(T)$, defined in
  Eq.~\ref{alp_Tdef}, as function of temperature in $2$-flavor QCD (filled
  symbols) and pure gauge theory (open symbols) from \cite{Kacze04}. The lines
  are explained in the text.  }
\label{alp_qqT}
\end{figure}
We also analyzed the temperature dependence of the maximal value that
$\alpha_{qq}(r,T)$ at fixed temperature exhibits at a certain distance,
$r_{max}$, {\em i.e.} we identify a temperature dependent coupling,
$\tilde{\alpha}_{qq}(T)$, defined as
\begin{eqnarray}
\tilde{\alpha}_{qq}(T)&\equiv&\alpha_{qq}(r_{max},T)\;.\label{alp_Tdef} 
\end{eqnarray}
Values for $\tilde{\alpha}_{qq}(T)$ are also available in pure gauge theory
\cite{Kacze04} at temperatures above deconfinement \footnote{In pure gauge
  theory $r_{max}$ and $\tilde{\alpha}_{qq}(T)$ would be infinite below
  $T_c$.}.  Our results for $\tilde{\alpha}_{qq}(T)$ in $2$-flavor QCD and pure
gauge theory are shown in Fig.~\ref{alp_qqT} as function of temperature,
$T/T_c$. At temperatures above deconfinement we cannot identify significant
differences between the data from pure gauge and 2-flavor QCD\footnote{Note,
  however, the change in temperature scale from $T_c\simeq200$ MeV in full to
  $T_c\simeq270$ MeV in quenched QCD.}. Only at temperatures quite close but
above the phase transition small differences between full and quenched QCD
become visible in $\tilde{\alpha}_{qq}(T)$. Nonetheless, the value of
$\tilde{\alpha}_{qq}(T)$ drops from about $0.5$ at temperatures only moderately
larger than the transition temperature, $T\;\gsim\;1.2T_c$, to a value of about
$0.3$ at $2T_c$. This change in $\tilde{\alpha}_{qq}(T)$ with temperature
calculated in $2$-flavor QCD does not appear to be too dramatic and can indeed
be described by the $2$-loop perturbative coupling assuming vanishing quark
masses. Due to the ambiguity in setting the scale in perturbation theory we
performed a best fit analysis to fix this scale for the entire temperature
range, $1.2\;\lsim\;T/T_c\;\lsim\;2$. We find $T_c/\Lambda=0.43(1)$ with
$\mu=2\pi T$. This is shown by the solid line (fit) in Fig.~\ref{alp_qqT}
including the error band (dotted lines).

At temperatures in the vicinity and below the phase transition temperature,
$T\;\lsim\;1.2T_c$, the behavior of $\tilde{\alpha}_{qq}(T)$ is, however, quite
different from the perturbative logarithmic change with temperature. The values
for $\tilde{\alpha}_{qq}(T)$ rapidly grow here with decreasing temperature and
approach non-perturbative large values. This again shows that
$\alpha_{qq}(r,T)$ mimics the confinement part of the zero temperature force
still at relatively large distances and that this behavior sets in already at
temperatures close but above deconfinement.

\vspace{-0.2cm}
\section{Summary} 
Our analysis of the heavy quark potential and coupling in $2$-flavor QCD at
$T=0$ shows that deviations from the string picture set in at $r\;\lsim\;0.4$
fm. At distances smaller than $0.3$ fm also deviations from $V(r)$ obtained
from Wilson loops in quenched QCD \cite{Necco} become apparent. The logarithmic
running of the coupling will become a dominant feature in $V(r)$ only for
$r\;\lsim\;0.1$ fm. We demonstrated that the QCD coupling at finite temperature
indeed runs with distance and coincides with the zero temperature running
coupling at sufficiently small distances.  Remnants of the confinement part of
the QCD force may survive the deconfinement transition and could play an
important role for the discussion of non-perturbative aspects of quark
anti-quark interactions at temperatures moderately above $T_c$. A clear
separation of the different effects usually described by color screening
($T\;\gsim\;T_c$) and effects commonly attributed to string-breaking
($T\;\lsim\;T_c$) is difficult to establish at temperatures in the close
vicinity of the confinement deconfinement cross over. Similar findings were
recently reported in quenched QCD \cite{Kacze04,Kacze03b}. Further details on
our study can be found in Refs.~\cite{Kacze05/1,Kacze05/2,Kacze05b}.
\subsection*{Acknowledgments}
\vspace{-0.2cm} We thank the Bielefeld-Swansea collaboration for providing us
their configurations with special thanks to S. Ejiri. We would like to thank E.
Laermann and F. Karsch for many fruitful discussions. F.Z. thanks P. Petreczky
for his continuous support. This work has partly been supported by DFG under
grant FOR 339/2-1 and by BMBF under grant No.06BI102 and partly by contract
DE-AC02-98CH10886 with the U.S. Department of Energy. At an early stage of this
work F.Z. has been supported through a stipend of the DFG funded graduate
school GRK881.

\end{document}